\documentclass{ws-procs9x6}

\begin{document}

\title{Anisotropic Flow in Ultra-Relativistic Heavy-Ion Collisions}

\author{Markus~D.~Oldenburg\\
\lowercase{(for the} STAR \lowercase{collaboration})}

\address{Ernest Orlando Lawrence Berkeley National Laboratory\\
One Cyclotron Road, Berkeley, CA 94720, USA\\ 
E-mail: MDOldenburg@lbl.gov}

\maketitle

\abstracts{The analysis of anisotropic flow of particles created in
high energy heavy-ion collisions gives insight into the early stage of
these reactions. Measurements of directed flow $(v_1)$, elliptic flow
$(v_2)$ and flow of $4^\mathrm{th}$ and $6^{\rm th}$ order ($v_4$ and
$v_6$) are presented. While the study of $v_2$ for multi-strange
particles establishes partonic collectivity the results for higher
order anisotropies constrain the initial conditions of hydrodynamic
model calculations.}

\section{Heavy-ion collisions}
High-energy collisions of heavy ions provide a tool to study nuclear
matter under extreme conditions. In the collision of two relativistic
nuclei, a hot and dense system of partonic matter may form. During the
formation time (of the order of $\tau_0 \sim 1$\,fm/$c$), partons from
the colliding nuclei suffer hard scatterings. High-$p_T$ hadrons are
predominantly produced in these hard scatterings and, therefore, carry
information about this early stage. If the system is thermalized and
the relevant degrees of freedom are partonic then the state of matter
created is referred to as the quark-gluon plasma.  Due to a rapid
expansion the system cools and at a critical temperature $T_{\rm c}$
undergoes a phase transition to hadronic matter\cite{Karsch}. This
critical temperature $T_{\rm c}$ might be close to the temperature of
chemical freeze-out $T_{\rm ch}$, at which the last inelastic
scatterings occur and particle yields no longer evolve. Thus hadron
abundances provide information about the system at $T_{\rm ch}$ $(\sim
T_{\rm c})$. The system further expands and elastic scattering
ceases. The corresponding temperature is called the kinetic freeze-out
temperature $T_{\rm fo}$. After kinetic freeze-out the momenta of the
produced particles are fixed. Therefore the momentum distributions
reflect the properties of the system at $T_{\rm fo}$.

In summary one has to note that the discussed signals of the early and
most interesting stage of the system's evolution are carried by only a
tiny fraction of all produced particles, namely those with high
transverse momentum $p_T$.

\section{\label{flow_explanation}Anisotropic flow}
In non-central heavy-ion collisions the overlap region of the two
colliding nuclei is not azimuthally symmetric in the transverse
plane. This leads to a larger pressure gradient in the reaction plane
(defined by the incident beam direction and the impact parameter) than
out-of-plane. These pressure gradients transform the original spatial
anisotropy of the system (which is out-of-plane) in an in-plane
anisotropy in momentum space. During this evolution the spatial
anisotropy diminishes and the process of generating anisotropies in
momentum space quenches itself. The measurement of the final azimuthal
momentum anisotropy is therefore a signal which is carried by all
particles, points back to the early evolution times of the collision,
and provides information about the equation of state of the generated
type of matter.

The transverse momentum asymmetry is analyzed utilizing a Fourier
decomposition of the momentum distribution in each event with respect
to the reaction plane\cite{flow}. By doing this one extracts Fourier
coefficients of different order $n$: $v_n \equiv \langle\cos\left(n \cdot
\phi\right)\rangle$, which characterize the strength of different
orders of flow. $v_1$ and $v_2$ are called directed and elliptic flow,
respectively.

\section{The STAR experiment}
The STAR experiment\cite{STAR} at the Relativistic Heavy Ion Collider
(RHIC) measures charged hadrons over a wide range in pseudorapidity
$\eta$ and transverse momentum $p_T$. Its main detector is a large
TPC, sitting in a magnetic field of 0.5\,T. Within the acceptance of
$|\eta| < 1.0$ particles are identified by specific energy loss ${\rm
d}E/{\rm d}x$. The two Forward-TPCs (FTPCs) add the region of $2.5 <
|\eta| < 4.0$ to the overall acceptance. Other detector subsystems,
which were not used in the studies shown here, include a silicon
vertex tracker, an electromagnetic calorimeter and time-of-flight
patches.

All results presented here were obtained using ${\rm Au} + {\rm Au}$
collisions at a center of mass energy of $\sqrt{s_{NN}} = 200$\,GeV.

\section{Elliptic flow $v_2$}
\begin{figure}[ht]
\centerline{\epsfxsize=0.5\linewidth\epsfbox{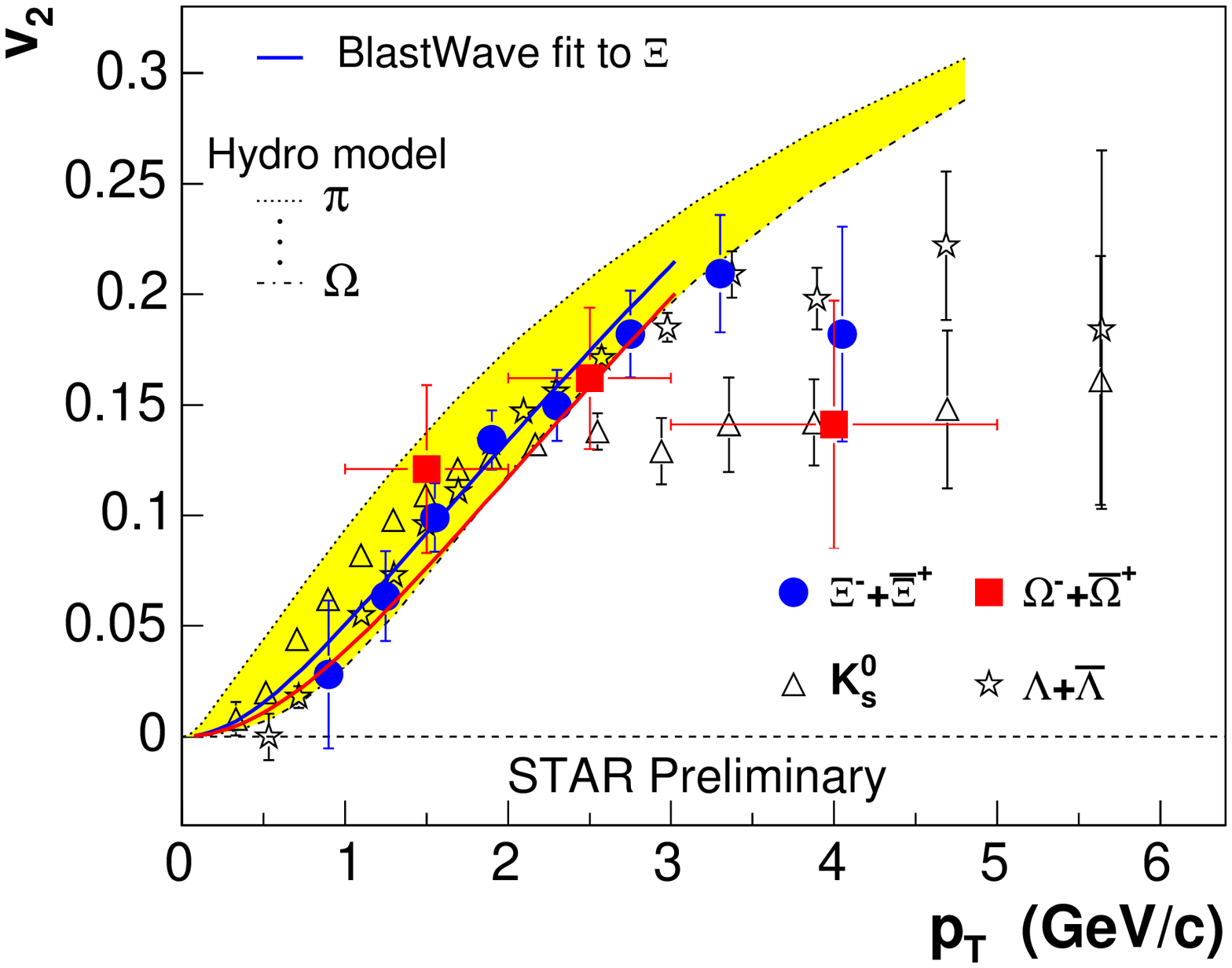}\hspace{0.2cm}\epsfxsize=0.48\linewidth\epsfbox{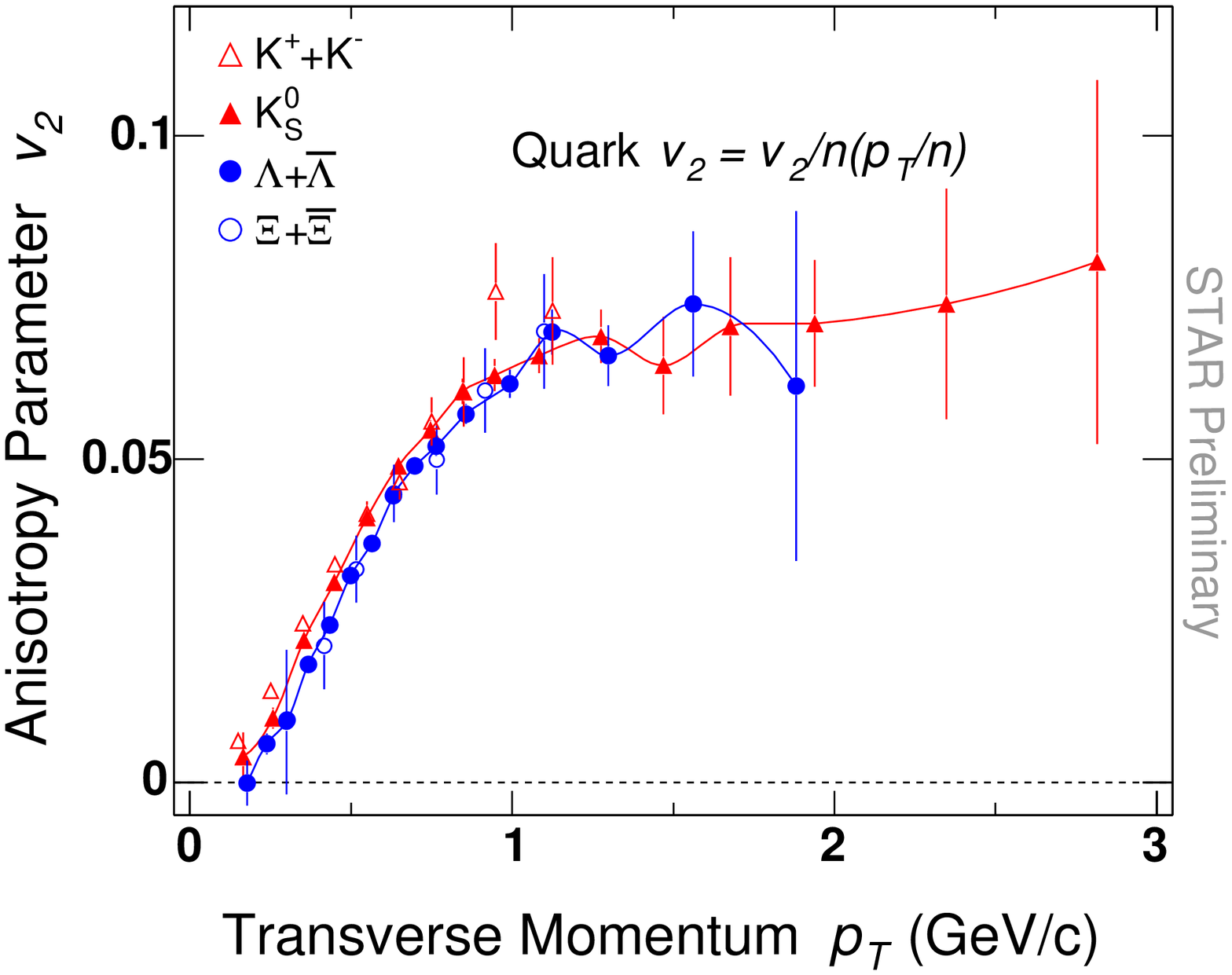}\hspace{-0.1cm}}   
\caption{\label{fig:v2}Elliptic flow $v_2$ as a function of transverse
  momentum $p_T$. The left panel shows $v_2$ for particles with
  different strangeness content, together with model predictions. On
  the right the measurements are rescaled by the number of constituent
  quarks.}
\end{figure}

In addition to the large overall elliptic flow $v_2$ of several
percent\cite{130GeVflow}, which suggests a strongly anisotropic
collective expansion, the measurement of $v_2$ shows a strong
dependence on $p_T$ (see Fig.\,\ref{fig:v2}, left). The almost linear
rise of $v_2(p_T)$ up to $p_T \sim 1.5$\,GeV/$c$ can for the first
time be reproduced by hydrodynamic models\cite{hydro}, which
overpredict elliptic flow at lower collision energies. These models
assume local thermal equilibrium. The mass dependence of $v_2(p_T)$ at
low $p_T$ agrees with these models as well, with lighter particles
experiencing more elliptic flow. This is consistent with all particles
flowing within the same velocity field, since heavier particles have
larger momentum for the same velocity.

STAR measured the elliptic flow $v_2(p_T)$ of the multi-strange
baryons $\Xi + \overline\Xi$ and $\Omega + \overline\Omega$ as
well\cite{multistrange}. Notably, their anisotropic flow is comparable
to the flow of the non-strange baryons. Since multi-strange baryons
are expected to have small hadronic cross sections, it is unlikely
that these particles could have picked up their large amount of flow
during the hadronic stage only. Therefore this independence of
strangeness content is a strong indication that collective motion is
established during the partonic stage already.

At intermediate $p_T$ ($1.5<p_T<5$\,GeV/$c$) a clear ordering of
mesons and baryons is visible. Quark coalescence models\cite{coal}
predict that $v_2$ scaled by the number of constituent quarks $n$ and
plotted as a function of $p_T/n$ will lie on top of each other. This
universal curve for mesons and baryons represents $v_2$ developed by
partons. As one can see from the measurement in Fig.\,\ref{fig:v2}
(right) quark coalescence in fact seems to be the dominant production
mechanism at intermediate $p_T$.

\section{Directed flow $v_1$ and the higher harmonics $v_4$ and $v_6$}
\begin{figure}[ht]
\centerline{\epsfxsize=0.5\linewidth\epsfbox{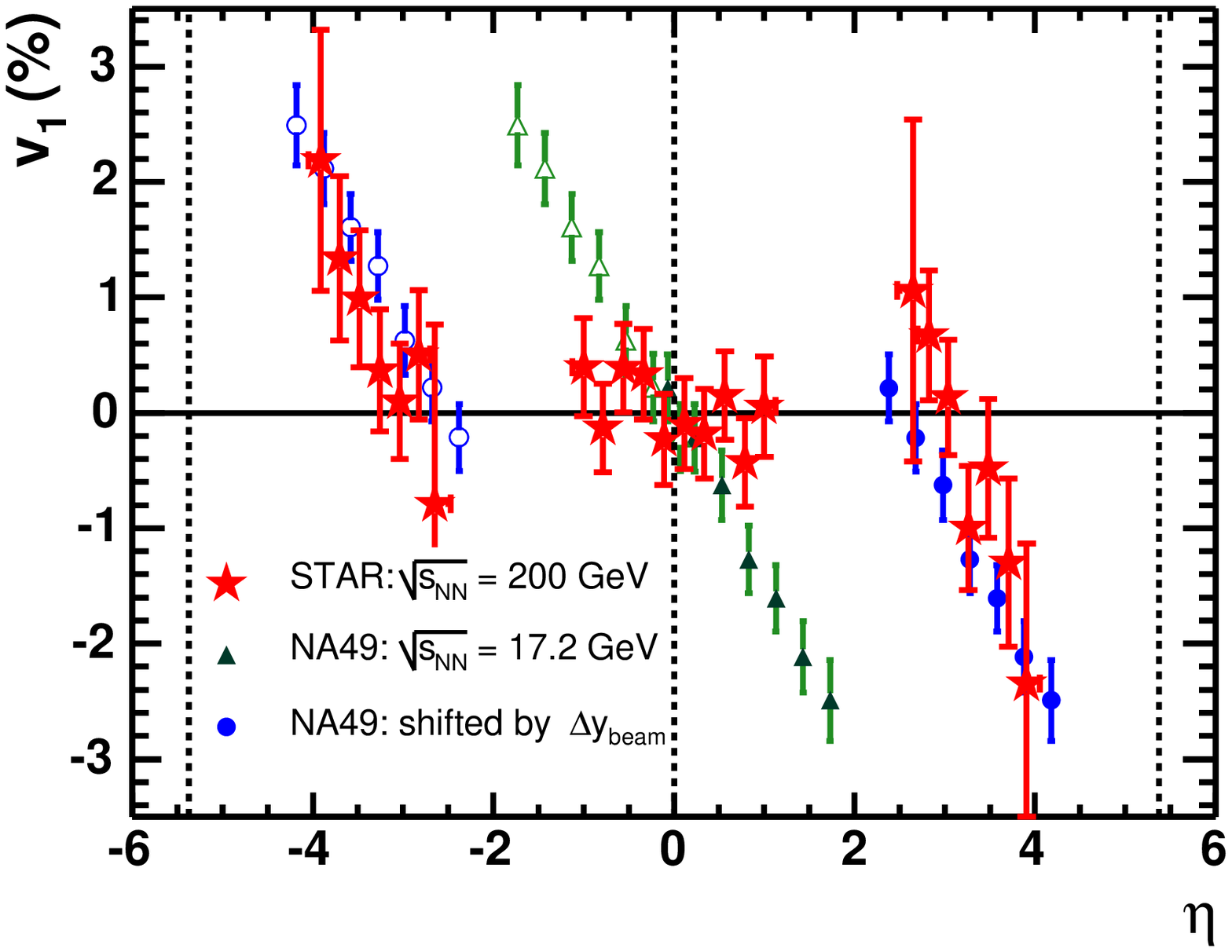}\hspace{0.5cm}\epsfxsize=0.51\linewidth\epsfbox{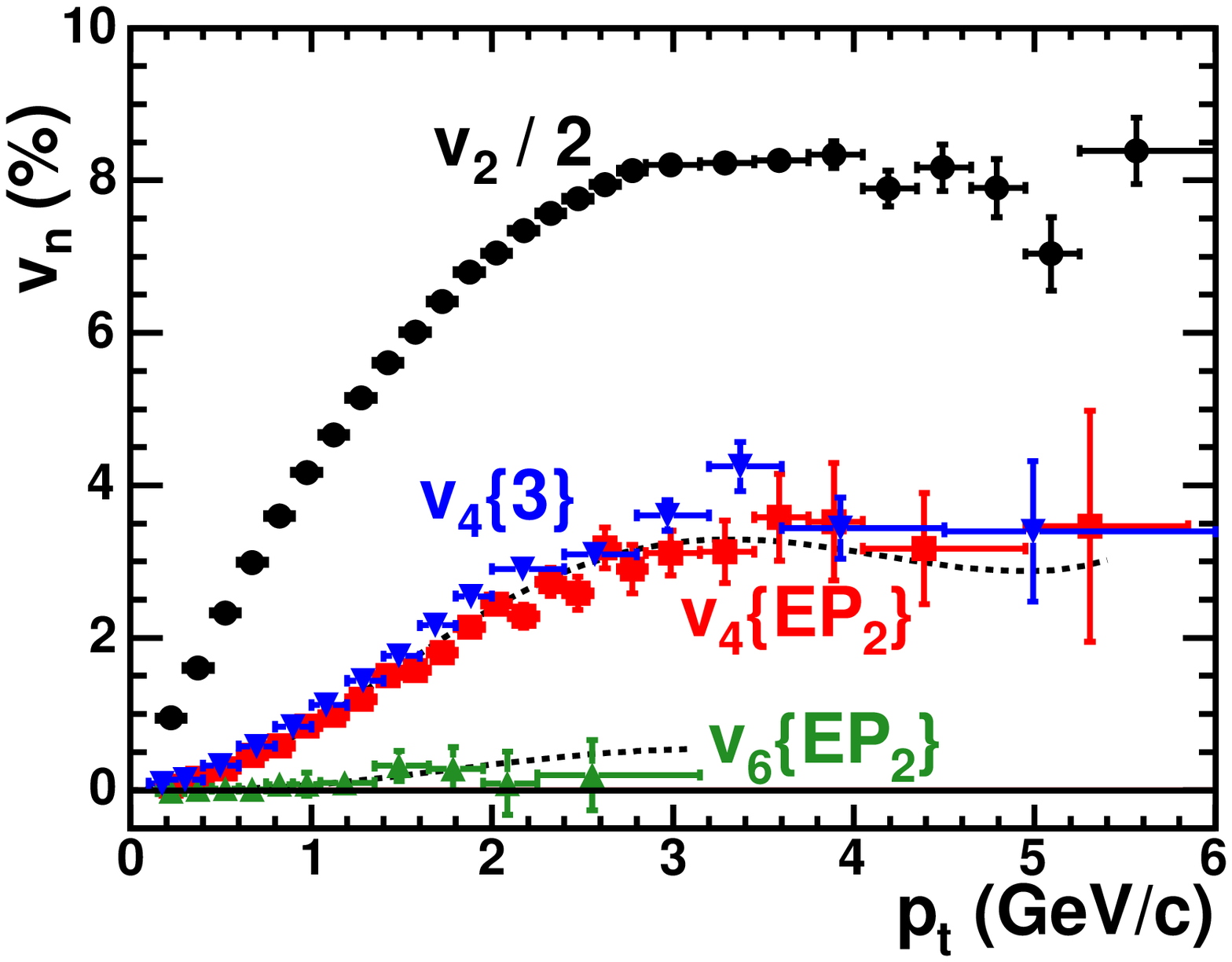}}   
\caption{\label{fig:vn}Measurements of anisotropic flow of different
orders. Left: Directed flow $v_1$ compared to results obtained by
NA49. Right: Anisotropic flow of $4^{\rm th}$ and $6^{\rm th}$ order
compared to the simple scaling of $v_2$ by $v_2^{n/2}$.}
\end{figure}

The measurement of directed flow $v_1$ proved to be difficult due to
its smallness at mid-rapidity. Therefore we utilized the
Forward-TPCs\cite{ftpc} to determine $v_1$. As can be seen in
Fig.\,\ref{fig:vn} on the left panel significant directed flow is only
visible in forward directions, while at mid-rapidity it is consistent
with zero\cite{v1v4}. Comparisons to results on $v_1$ at lower beam
energies look the same if plotted in the reference system of the
corresponding beam rapidity.

This measurement of directed flow $v_1$ was used to determine the sign
of elliptic flow $v_2$ which was still undetermined but assumed to be
positive\cite{130GeVflow}. The positiveness of $v_2$ could be
confirmed\cite{writeup} which is equivalent to the statement that
elliptic flow is
\emph{in-plane}. In other words the simple description given in
Sec.\,\ref{flow_explanation} holds and the initial spatial anisotropy
indeed transforms into a anisotropy in momentum space which is rotated
by $90^\circ$.

The strong elliptic flow made it possible to measure even higher
orders of transverse anisotropy. While we measure sizable flow of
$4^{\rm th}$ order, $v_4$, the signal of $v_6$ is consistent with zero
(see Fig.\,\ref{fig:vn}, right)\cite{v1v4}. These results help to better
constrain hydrodynamic model calculations, where the initial
conditions manifest themselves predominantly in the higher orders of
anisotropic flow\cite{Kolb}. Nevertheless, our measurements cannot be
explained by these models so far. As another surprising fact it was
noted that the higher harmonics of order $n$ scale like $v_2^{n/2}$.

\section{Conclusions}
In summary we have shown that the measurement of anisotropic flow of
different orders reveals interesting features of the early stage of
the two colliding gold nuclei. Directed flow $v_1$ confirmed that
elliptic flow $v_2$ is \emph{in-plane} and first measurements of flow
of higher orders give better constraints on hydrodynamic model
calculations. While elliptic flow at low transverse momentum
reproduces the mass scaling expected from hydrodynamic models, we
observe a meson-baryon scaling at intermediate $p_T$. The flow of
multi-strange baryons points to partonic collectivity.

\section*{Acknowledgments}
We thank the RHIC Operations Group and RCF at BNL, and the NERSC
Center at LBNL for their support. This work was supported in part by
the HENP Divisions of the Office of Science of the U.S. DOE; the
U.S. NSF; the BMBF of Germany; IN2P3, RA, RPL, and EMN of France;
EPSRC of the United Kingdom; FAPESP of Brazil; the Russian Ministry of
Science and Technology; the Ministry of Education and the NNSFC of
China; Grant Agency of the Czech Republic, FOM and UU of the
Netherlands, DAE, DST, and CSIR of the Government of India; the Swiss
NSF.

\appendix


\begin{thebibliography}{10}

\bibitem{Karsch} F.~Karsch, E.~Laermann, A.~Peikert, {\it Phys.~Lett.}~{\bf B478}, 447 (2000).

\bibitem{flow} S.~A.~Voloshin, Y.~Zhang, {\it Z.~Phys.}~{\bf C70}, 665 (1996).

\bibitem{STAR} K.~H.~Ackermann \emph{et al.}~(STAR collaboration), {\it
Nucl.~Instrum.~Meth.}~{\bf A499}, 624 (2003).

\bibitem{130GeVflow} K.~H.~Ackermann \emph{et al.}~(STAR
collaboration), {\it Phys.~Rev.~Lett.}~{\bf 86}, 402 (2001).

\bibitem{hydro} P.~Huovinen \emph{et al.}, {\it Phys.~Lett.}~{\bf
B503}, 58 (2001).

\bibitem{multistrange} J.~Castillo (for the STAR collaboration), to
appear in the proceedings of Quark Matter 2004, {\it J.~Phys.}~{\bf
G}, {\it arXiv:~nucl-ex/0403027} (2004).

\bibitem{coal} D.~Molnar, S.~A.~Voloshin, {\it Phys.~Rev.~Lett.}~{\bf
91}, 092301 (2003); V.~Grecco, C.~M.~Ko, P.~Levai, {\it Phys.~Rev.}~{\bf C68}, 034904 (2003); R.~J.~Fries, B.~M\"{u}ller, C.~Nomaka,
S.~A.~Bass, {\it Phys.~Rev.}~{\bf C68}, 044902 (2003); Z.~Lin,
C.~M.~Ko, {\it Phys.~Rev.~Lett.}~{\bf 89}, 202302 (2002).

\bibitem{ftpc}  K.~H.~Ackermann \emph{et al.}~(STAR collaboration), {\it
Nucl.~Instrum.~Meth.}~{\bf A499}, 713 (2003).

\bibitem{v1v4} J.~Adams \emph{et al.}~(STAR collaboration), {\it Phys.~Rev.~Lett.}~{\bf 92},
062301 (2004).

\bibitem{writeup} M.~D.~Oldenburg (for the STAR collaboration), {\it
arXiv:~nucl-ex/0403007} (2004).

\bibitem{Kolb} P.~F.~Kolb, {\it Phys.~Rev.}~{\bf C68}, 031902(R) (2003).

\end{thebibliography}
\end{document}